\documentclass[12pt]{article}
\pdfoutput=1
\usepackage{amssymb, amsmath,amsfonts}
\usepackage{multirow}
\usepackage{mathrsfs}
\usepackage{array}
\usepackage{cite}
\usepackage{booktabs}
\usepackage{tikz}
\usepackage{pgfplots}
\usepackage{float}
\usepackage{subfigure, graphicx}
\usepackage[pdftex, bookmarks=true,colorlinks,linkcolor=red,urlcolor=blue,citecolor=blue]{hyperref}
\usepackage{cite}
\usepackage{enumerate}
\usepackage{slashed}
\usepackage{tabularx}
\usepackage{enumerate} 
\usepackage{bm}
\usepackage{tikz}

\makeatletter\@addtoreset{equation}{section}\makeatother

\def\be{\begin{equation}}
\def\ee{\end{equation}}
\def\bea{\begin{eqnarray}}
\def\eea{\end{eqnarray}}
\newcommand{\comment}[1]{{\bf {\textcolor{blue}{ [#1]}}}}
\newcommand{\nn}{\nonumber}

\def\Dslash{\,\,{\raise.15ex\hbox{/}\mkern-12mu D}}
\def\Dbarslash{\,\,{\raise.15ex\hbox{/}\mkern-12mu {\bar D}}}
\def\delslash{\,\,{\raise.15ex\hbox{/}\mkern-9mu \partial}}
\def\delbarslash{\,\,{\raise.15ex\hbox{/}\mkern-9mu {\bar\partial}}}
\def\pslash{\,\,{\raise.15ex\hbox{/}\mkern-9mu p}}
\def\calDslash{\,\,{\raise.15ex\hbox{/}\mkern-12mu {\cal D}}}

\makeatletter\@addtoreset{equation}{section}\makeatother

\renewcommand{\title}[1]{\vbox{\center\LARGE{#1}}\vspace{5mm}}
\renewcommand{\author}[1]{\vbox{\center#1}\vspace{5mm}}
\newcommand{\address}[1]{\vbox{\center\em#1}}

\def\arXiv#1{\href{http://arxiv.org/abs/#1}{arXiv:#1}}
\def\arXiv#1#2{\href{http://arxiv.org/abs/#1}{arXiv:#1}}

\def\doi#1#2{\href{http://doi.org/#1}{#2}}

\begin{document}

\unitlength = .8mm

\begin{titlepage}
\vspace{.5cm}
 
\begin{center}
\hfill 
\hfill \\
\vskip 1cm

\title{Energy transport in holographic junctions}
\vskip 0.5cm
{Yan Liu$^{\,a,b}$}\footnote{Email: {\tt yanliu@buaa.edu.cn}}, ~{Chuan-Yi Wang$^{\,a, b}$}\footnote{Email: {\tt by2230109@buaa.edu.cn}} 

\address{${}^{a}$Center for Gravitational Physics, Department of Space Science,\\ 
Beihang University, Beijing 100191, China}

\address{${}^{b}$Peng Huanwu Collaborative Center for Research and Education, \\Beihang University, Beijing 100191, China}

\end{center}
\vskip 1.5cm

\abstract{We study energy transport in a conformal junction connecting three 2D conformal field theories using the AdS/CFT correspondence. The holographic dual consists of three AdS$_3$ spacetimes joined along the worldsheet of a tensile string anchored at the junction. Within a specific range of string tension, where the bulk solution is uniquely determined, we find that all energy reflection and transmission coefficients vary monotonically with the string tension. Notably, the total energy transmission, which quantifies the energy flux from one CFT through the junction, is bounded above by the effective central charge associated with both the CFTs and the junction. Results for energy transport  in conformal interfaces are recovered in a special limit. Furthermore, we extend our analysis to junctions connecting $N$ 2D CFTs.
}
\vfill

\end{titlepage}

\begingroup 
\hypersetup{linkcolor=black}
\tableofcontents
\endgroup





\newpage
\section{Introduction}
Conformal interfaces are codimension-one objects that connect two conformal field theories (CFTs), breaking half of their  conformal symmetry. They play important roles in the study of quantum critical systems, ranging from D-brane systems in string theory to condensed matter systems. 
Intriguingly, the properties of these  interfaces, such as the energy transmission coefficient  ($c_{LR}$) and the effective central charge ($c_\text{eff}$) which governs entanglement across the interface, are constrained by the central charges of the connected CFTs ($c_L$ and $c_R$) they connect. Specially, these quantities satisfy the inequality $0\leq c_{LR}\leq c_\text{eff} \leq \text{min}(c_L, c_R)$ \cite{Karch:2024udk}. Another important physical quantity characterizing the interface is the interface entropy. Due to their rich physical properties, conformal interfaces have recently attracted significant research interest  \cite{Andrei:2018die}. 

In this work, we generalize one-dimensional conformal interfaces to one-dimensional conformal junctions, i.e., spatially zero dimensional objects that connect multiple CFTs living on half-lines. We first focus on Y-shaped junctions connecting three CFTs, and then generalize to junctions joined by an arbitrary number $N$ of CFTs.

Like conformal interfaces, junctions are expected to be characterized by several important physical quantities, though their dynamics are more involved. For example, if an energy flux is injected from one CFT, it may split into multiple transmitted fluxes upon crossing the junction, leading to multiple transmission coefficients and a reflection coefficient. 
In field theory, such junctions can be described by boundary states, and their energy transports have been studied in \cite{Kimura:2015nka}. 
Entanglement entropy for these system has been studied in \cite{Gutperle:2017enx}. Here, we focus on strongly coupled CFTs and analyze them using holography. 

We consider a class of junction CFTs described by a holographic model involving a tensile string. 
In this setup, the junction corresponds to a tensile string anchored at the junction point, while the multiple CFTs that the junction connected to are dual to different AdS$_3$ geometries that terminate at the string's worldsheet. We first consider a  junction connecting three CFTs with  distinct central charges. Within a specific tension range, we find that the bulk solution is unique. We then study the system's fluctuations to extract its energy transport. Previous holographic studies of energy transport have primarily focused on conformal interfaces 
\cite{Bachas:2020yxv,Bachas:2022etu, Baig:2023ahz, Baig:2024hfc, Gutperle:2024yiz,  Baig:2022cnb, Liu:2025gle}. 
For conformal junctions, we find that there are multiple energy transport channels, all of which vary monotonically with string tension in the considered parameter regime. We define a total transmission coefficient and explore its relationship with the effective central charge as well as the central charges of the three CFTs. This allows us to generalize the inequality proposed in \cite{Karch:2024udk} to  conformal junctions. Finally, we extend our holographic study to junctions with arbitrary $N$ CFTs. 

This paper is organized as follows. In section \ref{sec:ft}, we review the field theory setup for junctions with an  arbitrary $N$ CFTs and define the energy transport coefficients. In section \ref{sec-3junction}, we first construct the holographic model for a three-junction system, then analyze its energy transport coefficients and discuss their relation with the effective central charges. In section \ref{sec:n}, we derive the energy transport for systems with a conformal defect connecting an arbitrary number $N$ of CFTs. In section \ref{sec:cd} we summarize our main results and discuss  open questions. Appendix \ref{app:a} contains the results on the energy transport in AdS/ICFT$_2$. In appendix \ref{app:b} we show the calculation of junction entropy for AdS/ICFT$_3$ and its generalization to ICFT$_N$.

\section{Brief
review of field theory results}
\label{sec:ft}
In this section, we briefly review the  field theory setup of the system and define the energy transports. The field theory framework was first developed in detail in \cite{Kimura:2015nka} and here we review the main results.  

\begin{figure}[h]
\begin{center}

\begin{tikzpicture}

\draw[very thick,color=blue] plot coordinates {(0,0)(2,0)};
\draw[very thick,color=black] plot coordinates {(2,0)(4,0)};
\draw [fill,color=red] (2,0) circle [radius=0.07];

\draw[very thick,color=blue] plot coordinates {(6,0)(8,0)};
\draw[very thick,color=orange] plot coordinates {(8,0)(8,2)};
\draw[very thick,color=black] plot coordinates {(8,0)(8,-2)};
\draw[very thick,color=gray] plot coordinates {(8,0)(6.586,1.414)};
\draw[very thick,color=magenta ] plot coordinates {(8,0)(6.586,-1.414)};
\draw[very thick,color=cyan] plot coordinates {(8,0)(9.414,1.414)};
\draw [fill,color=red] (8,0) circle [radius=0.07];
\draw[purple, thick,dashed] (8.2,-1.5) arc (-80:35:1.5cm);

\end{tikzpicture}

\end{center}
\vspace{-0.3cm}
\caption{\small Cartoon plot for ICFT$_2$ ({\em left}) and  ICFT$_N$ ({\em right}) at a constant time slice. The red dot is the location of defect and each CFT is defined on a half-line.}
\label{fig:ICFT2 and ICFTN}
\end{figure}
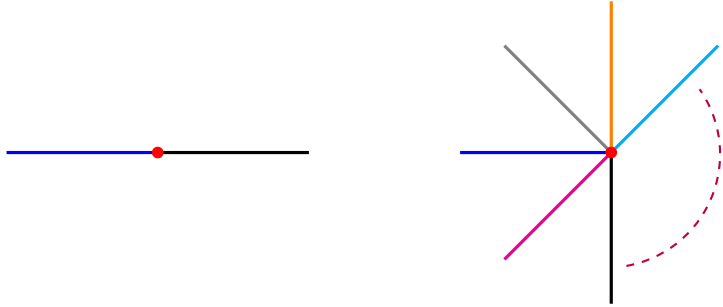

The system under consideration consists of an $N$-sheeted CFT joined along the worldline of a junction, which we denote as ICFT$_N$. This generalizes the standard  conformal interface ICFT$_2$, as illustrated in Fig. \ref{fig:ICFT2 and ICFTN}. The left panel shows ICFT$_2$, while the right panel is ICFT$_N$ at a constant time slice.  

Using the folding trick, the theory can be studied 
in terms of boundary state within the  folded theory, i.e. 
 CFT$_1\times\,$ $\text{CFT}_2\times\,\cdots\times\,{\text{CFT}}_N$.
\footnote{Note that each CFT$_A$ is  defined on the region $x_A<0$ for  $A=1,\cdots, N$.}
 The energy transmission and reflection coefficients can be defined through this boundary CFT framework.  
The boundary state $|b\rangle$ encodes  the boundary condition of the folded theory and satisfies the gluing condition \cite{Kimura:2015nka}, which generalizes earlier work \cite{Quella:2006de},   
\be
\label{eq:glu}
(L_n^\text{t}-\bar{L}_{-n}^\text{t})|b\rangle=0\,,
\ee
where $L_n^\text{t}=\sum_{\text{A}=1}^N L_n^\text{A}$ and $n\in {\bf Z}$. $L_n^\text{A}$ and $\bar{L}_n^\text{A}$ are the left and right Virasoro generators of CFT$_n$
. 

We define the R-matrix as
\be
R_\text{AB}=\frac{\langle 0| L^\text{A}_2\bar{L}^\text{B}_2| b\rangle}{\langle 0| b\rangle} \,.
\ee
Then we have 
\be
\mathcal{R}_\text{A}=\frac{2}{c_\text{A}} R_\text{AA}\,~~
\text{and}~~
\mathcal{T}_\text{AB}=\frac{2}{c_\text{A}} R_\text{AB}\,~{\text{with}~} \text{B}\neq \text{A}\,,~~~
\ee
where $\mathcal{R}_\text{A}$ is the energy reflection coefficient for CFT$_\text{A}$ and $\mathcal{T}_\text{AB}$ is the energy  transmission coefficient for transport from CFT$_\text{A}$ to CFT$_\text{B}$. 
Note that in general $\mathcal{T}_\text{AB}\neq \mathcal{T}_\text{BA}$. 

The gluing condition \eqref{eq:glu} imposes the following constraints 
\be\label{eq:rel1}
\sum_{\text{A}=1}^{N} R_\text{AB}=\frac{c_\text{B}}{2}\,,~~~
\sum_{\text{B}=1}^{N} R_\text{AB}=\frac{c_\text{A}}{2}
\ee
which lead to the relation  
\be\label{eq:rel2}
\mathcal{R}_\text{A}+\sum_{\text{B}\neq \text{A}} \mathcal{T}_\text{AB}=1\,.
\ee
We will verify these relations \eqref{eq:rel1} and \eqref{eq:rel2} through holographic calculations. 
Moreover, these constraints imply that the R-matrix contains $(N-1)^2$ independent parameters. 

We define $c_\text{AB}$ by the generalization of \cite{Meineri:2019ycm}
\be
\langle T_\text{A}(z_1) T_\text{B}(z_2)\rangle_\text{I}=\frac{c_\text{AB}/2}{(z_1-z_2)^4}
\ee
for $A\neq B$ and therefore $c_\text{AB}=c_\text{BA}$.  Note that the  expression above is in the unfolded picture.
The energy transports can also be defined through energy operator \cite{Meineri:2019ycm} and they satisfy  
\be
\mathcal{T}_\text{AB}=\frac{c_\text{AB}}{c_\text{A}}\,, 
\ee
or equivalently 
\be
\label{eq:rel3}
c_\text{A} \mathcal{T}_\text{AB}=c_\text{B} \mathcal{T}_\text{BA}\,.
\ee
From the average null energy condition (ANEC), we expect 
\be\label{eq:conscab}
0\leq c_\text{AB}\leq \text{min}(c_\text{A}, c_\text{B})\,,
\ee
or equivalently $\mathcal{T}_\text{AB}\in [0, \text{min}(1,\, c_\text{B}/c_\text{A})]$. 

\section{Holographic conformal junction connecting three CFTs}
\label{sec-3junction}

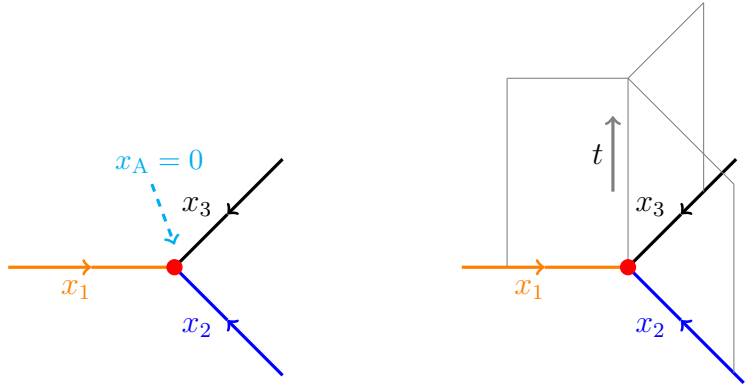
\begin{figure}[h!]
\begin{center}

\begin{tikzpicture}

\draw[very thick,color=orange] [->] (-2.2,0)--(-1.1,0);
\draw[very thick,color=orange]  (-1.1,0)--(0,0);
\node[color=orange] at (-1.3,-0.3) {$x_1$};
\draw[very thick,color=black] (0,0)--(0.7,0.7);
\draw[very thick,color=black] [->] (1.43,1.43)--(0.7,0.7);
\node[color=black] at (0.3,0.8) {$x_3$};
\draw[very thick,color=blue]  (0,0)--(0.7,-0.7);
\draw[very thick,color=blue] [->] (1.43,-1.43)--(0.7,-0.7);
\node[color=blue] at (0.3,-0.8) {$x_2$};
\draw [fill,color=red] (0,0) circle [radius=0.1];
\draw[very thick,color=orange] [->] (3.8,0)--(4.9,0);
\draw[very thick,color=orange]  (4.9,0)--(6,0);
\node[color=orange] at (4.7,-0.3) {$x_1$};
\draw[very thick,color=black] (6,0)--(6.7,0.7);
\draw[very thick,color=black] [->] (7.43,1.43)--(6.7,0.7);
\node[color=black] at (6.3,0.8) {$x_3$};
\draw[very thick,color=blue]  (6,0)--(6.7,-0.7);
\draw[very thick,color=blue] [->] (7.53,-1.53)--(6.7,-0.7);
\node[color=blue] at (6.3,-0.8) {$x_2$};
\draw[very thick,dashed, color=cyan] [->] (-0.3,1.1)--(0,0.3);
\node[color=cyan] at (-0.2,1.4) {
{\small $x_\text{A}=0$}};

\draw[color=gray]  (4.4,0)--(4.4,2.5);
\draw[color=gray]  (4.4,2.5)--(6,2.5);
\draw[color=gray]  (6,0)--(6,2.5);
\draw[color=gray]  (7,3.5)--(6,2.5);
\draw[color=gray]  (7,3.5)--(7,1);

\draw[color=gray]  (7.4,-1.4)--(7.4,1.1);
\draw[color=gray]  (6,2.5)--(7.4,1.1);

\draw[very thick,color=gray] [->] (5.8,1)--(5.8,2);
\node[very thick,color=black] at (5.6,1.5) {$t$};

\draw [fill,color=red] (6,0) circle [radius=0.1];

\end{tikzpicture}

\end{center}
\vspace{-0.3cm}
\caption{\small Cartoon plot for ICFT$_3$ at a constant time slice ({\em left}) and with time direction added ({\em right}). The red dot marks the defect location at $x_A=0$ and each CFT defined on the half-line $x_\text{A}<0,$ where $\text{A}=1,2,3$ .}
\label{fig:3CFT_and_sp_dual}
\end{figure}

In this section, we focus on the conformal junction connecting three CFTs. The field theory configuration is illustrated in Fig. \ref{fig:3CFT_and_sp_dual}. We will construct the explicit holographic dual and study the energy transports through this junction. 

The gravitational theory is described by the  following action,  
\be
\label{eq:action}
S_\text{bulk}=\sum_{\text{A}=1}^{3} S_\text{A} +S_Q 
\ee
where
\begin{align}
S_\text{A}&=\int_{N_\text{A}} d^3x\sqrt{-g_\text{A}}\,\bigg[\frac{1}{16\pi G}\bigg(R_\text{A}+\frac{2}{L_\text{A}^2}\bigg)
    \bigg] \,, \\
S_{Q}&=\frac{1}{8\pi G}\int_Q d^2y\sqrt{-h}\, \bigg[\big(K_\text{1}+K_\text{2}+K_\text{3}\big)-T\bigg]\,.
\end{align}
The coordinates on $N_\text{A}$ are $x_\text{A}^a=(u_\text{A}, t_\text{A}, x_\text{A})$. We choose the AdS boundary of $N_\text{A}$ to lie in the region $x_\text{A}<0$, while the boundary of the junction string $Q$ is located at $x_\text{A}=0$. 

The continuity condition for the metric on $Q$ is given by 
\be
\label{eq:continuous}
h_{\mu\nu}=
g_{ab}^{\text{A}}\frac{\partial x_\text{A}^a}{\partial y^\mu}
\frac{\partial x_\text{A}^b}{\partial y^\nu}\bigg{|}_\text{Q}\,  , ~~~~~(\text{A}=1,2,3)  \,  .
\ee
The equations of motion are
\begin{align}
R^{\text{A}}_{ab}-\frac{1}{2}g^{\text{A}}_{ab}R^{\text{A}}-\frac{1}{L^2_\text{A}}g^{\text{A}}_{ab}=&~0  \, 
,~~~~ (\text{A}=1,2,3)  \,  \label{eq:bulk}\\
\sum_{A=1}^3K_{\mu\nu}^{\text{A}}-\Big(\sum_{A=1}^3 K^{\text{A}}-T\Big)\,h_{\mu\nu}=&~0 \label{eq:q} \, .
\end{align}
Equation \eqref{eq:q} can be simplified as
\begin{align}
\sum_{\text{A}=1}^3K_{\mu\nu}^{\text{A}}-T h_{\mu\nu}=0  \, .
\end{align}
The junction conditions for multiple junctions have been previously studied in \cite{Shen:2024itl, Shen:2024dun}.

\subsection{Background solution}
The bulk geometry for $N_\text{A}$ is the planar AdS$_3$ solution
\begin{align}
\label{eq:ads3metric}
ds^2_\text{A}=\frac{L_\text{A}^2}{u_\text{A}^2}\bigg[-dt_\text{A}^2+
dx^2_\text{A}+du_\text{A}^2
\bigg]\,, ~~~ \text{A}=\text{1,~2,~3}\,.
\end{align}
The junction string is given by
\begin{align}
x_\text{A}=u_\text{A} \tan\theta_\text{A}  \,  .
\end{align}

\begin{figure}[h]
\begin{center}

\begin{tikzpicture}
\draw[very thick] plot coordinates {(-3,0)(0,0)};
\node at (-2.5,-0.3) {CFT$_\text{A}$};

\node at (-1.5,2) {$(N_\text{A},g^\text{A}_{ab})$};

\draw[color=red] [->] (0,0) -- (3,0);
\draw[color=red]  [->] (0,0) -- (0,3);
\node[color=red] at (3,-0.3) {$x_\text{A}$};
\node[color=red] at (0.3,3) {$u_\text{A}$};

\draw[color=blue][very thick] [-] (0,0) -- (1,3);

\draw[color=black] [->] (0.9,2.1) -- (1.2,3);
\draw[color=black] [->] (0.9,2.1) -- (1.8,1.8);

\node[color=black] at (1.6,3) {$w_\text{A}$};
\node[color=black] at (2.1,1.8) {$v_\text{A}$};

\node[color=black] at (0.25,1.7) {$\theta_\text{A}$};

\end{tikzpicture}

\end{center}
\vspace{-0.3cm}
\caption{\small The plot for CFT$_A$ and its bulk dual geometry $N_\text{A}$. The junction brane $Q$ is shown in blue. The AdS boundary is at $u_\text{A}\to 0$. The entire system is formed by gluing $N$ such subsystems via maps that satisfy specific  gluing conditions. 
}
\label{fig:dualCFT}
\end{figure}
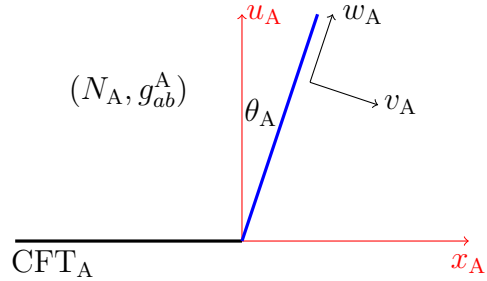

Note that $\theta_A$ parameterizes the angle between the brane and the perpendicular direction, see Fig. \ref{fig:dualCFT}. We consider the case 
\begin{align}
-\frac{\pi}{2} <\theta_\text{A}<\frac{\pi}{2} \,,
\end{align}
where $\theta_\text{A}$ is positive when the brane lies on the right half of the spacetime, as illustrated in Fig. \ref{fig:dualCFT}. 

For later convenience, we introduce rotated coordinates $(v_\text{A},t_\text{A},w_\text{A})$,
which are related to the Poincare coordinates in \eqref{eq:ads3metric} via 
\begin{align}
v_\text{A}= x_\text{A}\cos\theta_\text{A}-
u_\text{A} \sin\theta_\text{A}   \,  ,~~~
w_\text{A}= x_\text{A}\sin\theta_\text{A}+
u_\text{A} \cos\theta_\text{A}  \, .
\end{align}
In these coordinates, the brane is located at $v_\text{A}=0$, and the induced metric on $Q$ becomes
\begin{align}
ds^2_{Q}=\frac{L^2_\text{A}}
{(w_\text{A} \cos\theta_\text{A})^2}\,
(-dt^2_\text{A}+dw^2_\text{A})  \, .\label{eq:brane_geo}
\end{align}
The region where CFT$_\text{A}$ lives corresponds to 
\begin{align}
v_\text{A} \le 0 \, ,\ \ 
w_\text{A}=
v_\text{A} \tan\theta_\text{A} \, ,
\end{align}
which is equivalent to 
$
x_\text{A}\le 0 \, ,
u_\text{A}=0 \, .
$

Thus the continuity condition \eqref{eq:continuous} for the metric on $Q$ can be solved by requiring
\begin{align}
& t_\text{1}=t_\text{2}=t_\text{3}\equiv t \, ,~~~
w_\text{1}=w_\text{2}=w_\text{3}\equiv z \, , ~~~
\frac{L_\text{1}}{\cos\theta_{1}}=
\frac{L_\text{2}}{\cos\theta_{2}}=
\frac{L_\text{3}}{\cos\theta_{3}} \equiv L_a \label{eq:bg jc} \,  ,
\end{align}
where we use $(t,z)$ as the coordinates on $Q$. From \eqref{eq:brane_geo}, we observe that $L_a$ is the AdS radius of the brane geometry. 

The central charges for the dual CFTs are given by 
\be
c_\text{A}=\frac{3L_\text{A}}{2G} \, 
~~~\text{with~} A\in (1, 2, 3)
\ee
as first derived in \cite{Brown:1986nw, Bachas:2001hpy}. We further  define
\begin{align}
\label{eq:ca}
c_a=\frac{3L_a}{2G} \, ,
\end{align}
where $L_a$ 
is specified in \eqref{eq:bg jc}.

The junction condition \eqref{eq:q} is solved by
\begin{align}
\label{eq:relT}
T=\sum_{\text{A}=1}^3 \frac{\sin\theta_\text{A}}{L_\text{A}}  \,  .
\end{align}

For a system with a fixed tension $T$ and the AdS radii $L_1, L_2, L_3$, one can solve the system  from \eqref{eq:bg jc} and \eqref{eq:relT} to obtain the bulk configuration. 

\subsubsection{Range of tension}
\label{subsec:rangetension}

In this subsection, we constrain the allowed values of the brane tension by requiring that the bulk solution is uniquely determined for a given  tension $T$.  

First, the null energy condition (NEC) imposes  $T\geq 0$.\footnote{We thank Policastro Giuseppe for discussions on this point.} To see this, consider the stress tension on the junction brane 
$ T^{Q}_{\mu\nu}=-T h_{\mu\nu}  \, , $
or equivalently 
$
T^{Q}_{ab}=-T (g_{ab}-n_{a}n_{b} ) \, ,
$
where $n^{a}$ is spacelike normal vector to the brane. For a null vector $k^{a}$ is of the form 
$k^{a}=t^{a}+n^{a} \, ,$ where 
$t^{a}$ is a timelike vector tangent to the brane, the NEC becomes
\begin{align}
T^{Q}_{ab}k^{a}k^{b}=
T\ge 0
\, .
\end{align}

\begin{figure}[h!]
\begin{center}
\includegraphics[width=0.46\textwidth]{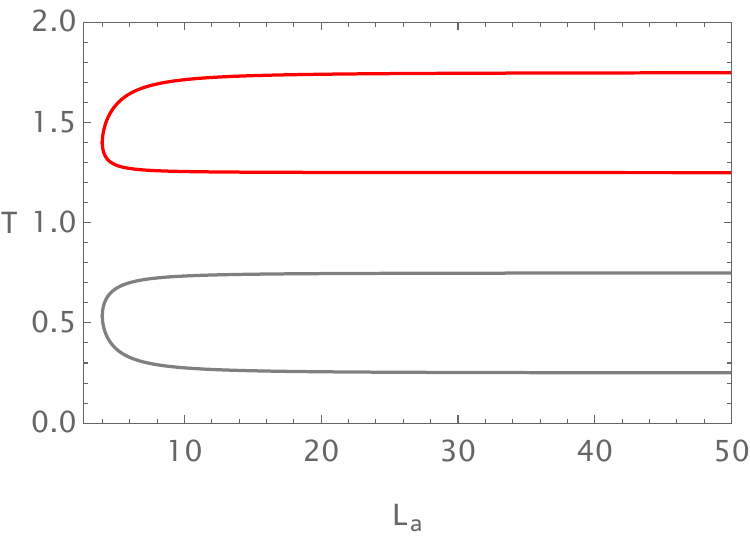}
~~~~~
\includegraphics[width=0.46\textwidth]{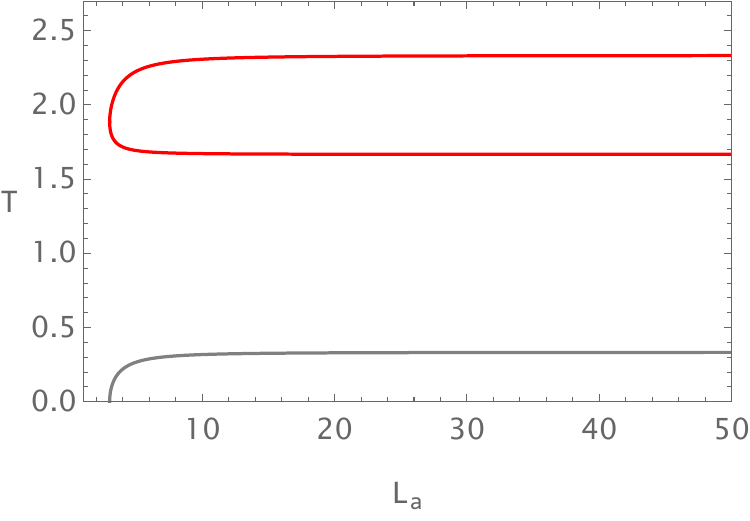}
\\
\includegraphics[width=0.46\textwidth]{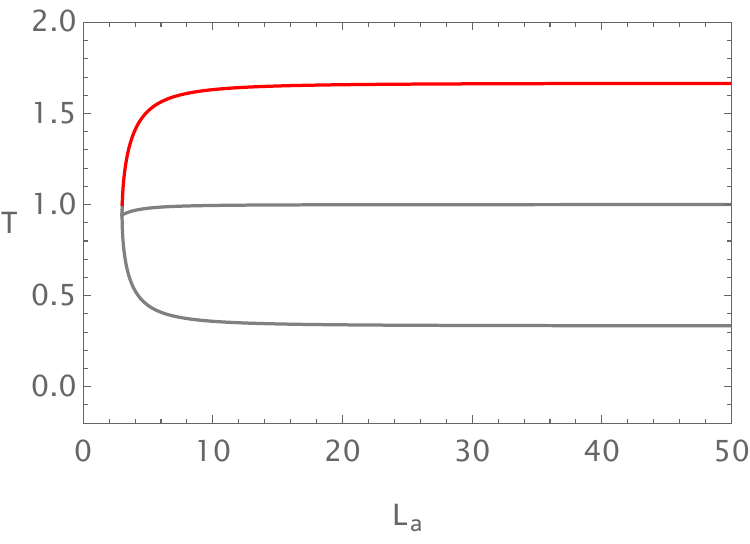}
~~~~~
\includegraphics[width=0.46\textwidth]{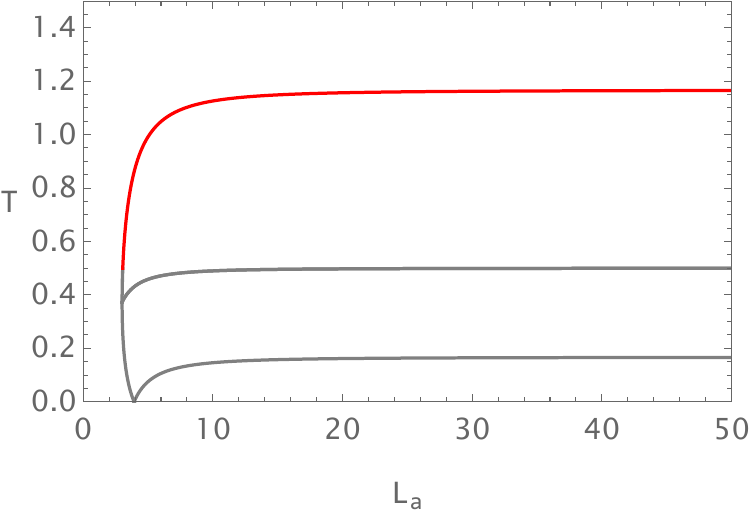}
\end{center}
\vspace{-0.3cm}
\caption{\small  The plots of tension $T$ as a function of $L_a$ with different AdS radii  
$L_1=1,L_2=2,L_3=4$ ({\em top-left}),
$L_1= L_2=1,L_3=3$ ({\em top-right}),
$L_1=1,L_2=L_3=3$ ({\em bottom-left})
$L_1=2,L_2=L_3=3$ ({\em bottom-right}).  
The red lines indicate the tension ranges specified in \eqref{eq:tensionran1} and \eqref{eq:tensionran2}, which are considered for energy transport calculations. 
}
\label{fig:tension}
\end{figure}

Second, the upper bound on the tension can be easily obtained analytically as 
\begin{align}
 T\,<\, \sum_{\text{A}=1}^3\frac{1}{L_\text{A}}
 \,  .
\end{align}
We focus on the regime of positive tension. Interestingly, the allowed tension range varies depending on the central charges of the CFTs. Our numerical results for the tension as a function of $L_a$ are shown in Fig. \ref{fig:tension}, from which we see that the tension range exhibits more  gaps compared to the ICFT case (see the tension range analysis for ICFT in appendix \ref{app:a}).

For the case where $L_1\leq L_2<L_3$ and $L_1+L_2<L_3$,
\footnote{
Note that when $L_1\leq L_2<L_3
$, the condition $L_1+L_2<L_3$ is sufficient to guarantee the  existence of a unique bulk  solution. Numerically, we have verified through several examples satisfying $L_1+L_2>L_3$ that multiple bulk configurations exist for a given value of $T$. 
Note that the Euclidean on-shell action for the zero temperature solution is zero.
}
the range of positive tension further depends on the sign of $\frac{1}{L_1}-\frac{1}{L_2}-\frac{1}{L_3}$. When $\frac{1}{L_1}-\frac{1}{L_2}-\frac{1}{L_3}\ge 0$,
the positive tension range is
\be\label{eq:tenran1}
T\in 
\left( 
\frac{1}{L_1}-\frac{1}{L_2}-\frac{1}{L_3}\,,~ \frac{1}{L_1}-\frac{1}{L_2}+\frac{1}{L_3}
\right)
\bigcup  
\left( 
\frac{1}{L_1}+\frac{1}{L_2}-\frac{1}{L_3}\,,~ \frac{1}{L_1}+\frac{1}{L_2}+\frac{1}{L_3}
\right)  \, ,
\ee
where the second open interval corresponds to the red curve in the top-left picture in Fig. \ref{fig:tension}. 
On the other hand, when 
$\frac{1}{L_1}-\frac{1}{L_2}-\frac{1}{L_3}< 0$,
the positive tension range becomes
\be\label{eq:tenran2}
T\in 
\left[ 
0\,,~ \frac{1}{L_1}-\frac{1}{L_2}+\frac{1}{L_3}
\right)
\bigcup  
\left( 
\frac{1}{L_1}+\frac{1}{L_2}-\frac{1}{L_3}\,,~ \frac{1}{L_1}+\frac{1}{L_2}+\frac{1}{L_3}
\right)  \, ,
\ee
where the second open interval is represented by the red curve in the top-right picture in Fig. \ref{fig:tension}. 

We can prove analytically that 
for any given value of tension within these  regimes, there exists a unique bulk configuration. 
For simplicity, we focus on the region smoothly connected to the largest tension  
\be\label{eq:tensionran1}
T\in \left( 
\frac{1}{L_1}+\frac{1}{L_2}-\frac{1}{L_3}\,,~ \frac{1}{L_1}+\frac{1}{L_2}+\frac{1}{L_3}
\right)\,. 
\ee
Furthermore, we will show that in this range \eqref{eq:tensionran1} with $L_1\leq L_2<L_3$ and  $L_1+L_2<L_3$, the junction entropy spans from $-\infty$ to $\infty$. Another motivation for  excluding the first region in \eqref{eq:tenran1} and \eqref{eq:tenran2} comes from AdS/ICFT$_2$, where the lower tension bound corresponds to the Coleman-De lucia vacuum decay bound \cite{Bachas:2020yxv, Baig:2022cnb}. This suggests that in AdS/ICFT$_3$ the first tension region in \eqref{eq:tenran1} and \eqref{eq:tenran2} might similarly be unstable under the vacuum decay \cite{Maldacena:2010un}. 

For the case where $L_1\leq L_2=L_3$, the region of positive tension depends on the sign of $L_1-\frac{L_2}{2}$. When $L_1>\frac{L_2}{2}$, the tension range is 
\be
T\in \left[ 
0\, ,
\frac{1}{L_1}+\frac{2}{L_2}
\right)
\, .
\ee 
When $L_1\le \frac{L_2}{2}$, the tension range becomes
\be
T\in 
\left(
\frac{1}{L_1}-\frac{2}{L_2}\, ,
\frac{1}{L_1}+\frac{2}{L_2}
\right)
\, .
\ee 
To ensure a unique bulk configuration for a given tension, we focus on the region 
\be \label{eq:tensionran2}
T\in 
\left[ 
\frac{1}{L_1}\, ,
\frac{1}{L_1}+\frac{2}{L_2}
\right)\, .
\ee
The red curves in the bottom-left and bottom-right panels of Fig. \ref{fig:tension} correspond to this regime.

To prevent the holographic system from consisting of a superposition states of two or more classical spacetimes for given tension values and dual CFT central charges, we restrict our discussion to the two cases and a specified tension ranges \eqref{eq:tensionran1} and \eqref{eq:tensionran2}.

\subsection{Calculation of energy transport}
To compute the energy transport coefficients, we consider first-order perturbations of system using the method developed in  \cite{Bachas:2020yxv}. The perturbed bulk metric on each sector (\text{A}=\text{1,~2,~3}) takes the form  
\be
ds^2=ds^2_\text{A}+[ds^2]^{(2)}_\text{A}\,,~~
\ee
where  
\begin{align}
\begin{split}
\label{eq:pertur}
[ds^2]^{(2)}_{1}=&\,4GL_{1}\epsilon
\left[\,
e^{i\omega (t_1-x_1)}(dt_1-dx_1)^2+
\mathcal{R}_{1}e^{i\omega (t_1+x_1)}(dt_1+dx_1)^2
\right] +c.c. \,  , \\
[ds^2]^{(2)}_{2}=&\,4GL_{2}\epsilon\,
\mathcal{T}_{1 2}e^{i\omega (t_2+x_2)}(dt_2+dx_2)^2
 +c.c. \,  , \\
 [ds^2]^{(2)}_{3}=&\,4GL_{3}\epsilon\,
\mathcal{T}_{1 3}e^{i\omega (t_3+x_3)}(dt_3+dx_3)^2
 +c.c. \,  ,
 \end{split}
\end{align}
where coefficients $\mathcal{R}_{1}$, $\mathcal{T}_{12}$ and $\mathcal{T}_{13}$ characterizing the energy reflection and transmission respectively.

The perturbed brane is parameterized as
\begin{align}
t_\text{A}=t+4G\epsilon e^{i\omega t}\lambda_\text{A}(z) \,  ,\ \ 
v_\text{A}=4G\epsilon e^{i\omega t}\delta_\text{A}(z)
\,  ,\ \ 
w_\text{A}=z+4G\epsilon e^{i\omega t}\xi_\text{A}(z) \, ,\ 
(\text{A}=1,2,3)  \,  ,
\end{align}
from which we find the equation for the junction  brane $
v_\text{A}-4G\epsilon e^{i\omega t_\text{A}}\delta_\text{A}(w_\text{A})=0
\, .
$

We define the following quantities 
\begin{align}
\begin{split}
\bm{I}=e^{-i\omega\sin\theta_1 z}\, ,&~~~~~~
\bm{R}_1=e^{+i\omega\sin\theta_1 z}  \, ,\\
\bm{T}_{1 2}=\mathcal{T}_{1 2}e^{+i\omega\sin\theta_2 z}\, ,&~~~~~~
\bm{T}_{1 3}=\mathcal{T}_{1 3}e^{+i\omega\sin\theta_3 z}
\end{split}
\end{align}
and the variables 
\begin{align}
\label{eq:par-firstorder}
\begin{split}
&\xi_{12}=\xi_1-\xi_2\, ,\ \ 
\lambda_{12}=\lambda_1-\lambda_2\, ,\ \ 
\delta_{12}=\tan\theta_1\,\delta_1-\tan\theta_2\,\delta_2 \, , \\
&\xi_{32}=\xi_3-\xi_2\, ,\ \ 
\lambda_{32}=\lambda_3-\lambda_2\, ,\ \ 
\delta_{32}=\tan\theta_3\,\delta_3-\tan\theta_2\,\delta_2 \, .
\end{split}
\end{align}
The system is characterized by seven independent physical quantities: $\delta_2$ and the six variables in \eqref{eq:par-firstorder}. 

The continuity condition of the induced metric between the first and second bulk regions 
\be
g_{ab}^\text{A}\frac{\partial x_\text{A}^a}{\partial y^\mu}
\frac{\partial x_\text{A}^b}{\partial y^\nu}\bigg{|}_\text{A=1}
=\,g_{ab}^\text{A}\frac{\partial x_\text{A}^a}{\partial y^\mu}
\frac{\partial x_\text{A}^b}{\partial y^\nu}\bigg{|}_\text{A=2}  \, 
\ee
gives three equations
\begin{align}
\begin{split}
\delta_{12}-\xi_{12}+i\omega z\lambda_{12}
=&\,
\frac{z^3}{2L_a}
\left[\, 
 (\bm{I}+\bm{R}_1)\cos\theta_1-
 \bm{T}_{1 2}\cos\theta_2
\right]  \, ,\\
i\omega\xi_{12}-\lambda'_{12}
=&\,
\frac{z^2}{2L_a}
\left[\, 
(\bm{I}-\bm{R}_1)\sin (2\theta_1)
+\bm{T}_{1 2} \sin (2\theta_2)
\right]   \, ,\label{eq:1st-2}\\
\delta_{12}-\xi_{12}+z\xi'_{12}
=&\,
\frac{z^3}{2L_a}
\left[ 
-(\bm{I}+\bm{R}_1) (\sin\theta_1)^2\cos\theta_1
+\bm{T}_{1 2}
(\sin\theta_2)^2\cos\theta_2
\right]  \, , 
\end{split}
\end{align}
where primes denote $z$-derivatives. 
The solutions of 
these three equations 
are
\begin{align}
\begin{split}
\lambda_{12}=&\,
\frac{
i(2-\omega^2 z^2 (\cos\theta_1)^2)
}{
2\omega^3 L_a\cos\theta_1
}\,
(\bm{I}+\bm{R}_1)
-
\frac{
i(2-\omega^2 z^2 (\cos\theta_2)^2)
}{
2\omega^3 L_a\cos\theta_2
}\,\bm{T}_{1 2}
+b_1 e^{-i\omega z} \, , \\
\delta_{12}=&\,
-\frac{
i
(2+\omega^2 z^2 (\cos\theta_1)^2 + 2i\omega z\sin\theta_1) \tan\theta_1
}{
2\omega^3 L_a
}\, \bm{I}   \, \\
&+\frac{
i
(2+\omega^2 z^2 (\cos\theta_1)^2 - 2i\omega z\sin\theta_1) \tan\theta_1
}{
2\omega^3 L_a
} \, \bm{R}_1   \,  \\
&-\frac{
i
(2+\omega^2 z^2 (\cos\theta_2)^2 - 2i\omega z\sin\theta_2) \tan\theta_2
}{
2\omega^3 L_a
}\, \bm{T}_{1 2}
-b_1 (1+i\omega z)e^{-i\omega z} \, ,\\
\xi_{12}=&\,
\frac{
i(2i\omega z\cos\theta_1
-\omega^2 z^2\cos\theta_1 \sin\theta_1
-2\tan\theta_1
)
}{
2\omega^3 L_a
} \,\bm{I}   \, \\
&+
\frac{
i(2i\omega z\cos\theta_1
+\omega^2 z^2\cos\theta_1 \sin\theta_1
+2\tan\theta_1
)
}{
2\omega^3 L_a
} \,\bm{R}_1  \, \\
&+
\frac{
2\omega z\cos\theta_2
-i\omega^2 z^2\cos\theta_2 \sin\theta_2
-2i\tan\theta_2
}{
2\omega^3 L_a
}\, \bm{T}_{1 2}
-b_1 e^{-i\omega z}
\, ,
\end{split}
\end{align}
where $b_1$ is an integration constant and we have used the infalling-wave boundary condition at IR. The sourceless conditions 
$\lambda_{12}(0)=\delta_{12}(0)=0$
give
\begin{align}
\begin{split}
\mathcal{R}_1=&\,
\frac{
-1+\sin\theta_1
}{
1+\sin\theta_1
}
+\mathcal{T}_{1 2}
\frac{\cos\theta_1}{1+\sin\theta_1}
\left(
\frac{\cos\theta_2}{1+\sin\theta_2}
\right)^{-1}  \, , \label{eq:R1T2} \\
b_1=&\,
\frac{i}{\omega^3 L_a (1+\sin\theta_1)}
\left[ 
-2\tan\theta_1
+\mathcal{T}_{1 2}
\left(
\frac{\sin\theta_1}{\cos\theta_2}
-\tan\theta_1
\right)
\right]   \, ,
\end{split}
\end{align}
and $\xi_{12}(0)=0$ can be derived from $\lambda_{12}(0)=\delta_{12}(0)=0$.

Similarly, from the continuity condition of the induced metric at first order between the third and second bulk regions
\be
g_{ab}^\text{A}\frac{\partial x_\text{A}^a}{\partial y^\mu}
\frac{\partial x_\text{A}^b}{\partial y^\nu}\bigg{|}_\text{A=3}
=\, g_{ab}^\text{A}\frac{\partial x_\text{A}^a}{\partial y^\mu}
\frac{\partial x_\text{A}^b}{\partial y^\nu}\bigg{|}_\text{A=2}  \,, 
\ee
three equations can be obtained 
\begin{align}
\label{eq:isreal2}
\begin{split} 
\delta_{32}-\xi_{32}+i\omega z\lambda_{32}
=&\,
\frac{z^3}{2L_a}\left[\, 
-\bm{T}_{1 2}\cos\theta_2
+\bm{T}_{1 3}\cos\theta_3
\right]   \, ,  \\
i\omega\xi_{32}-\lambda'_{32}
=&\,
\frac{z^2}{2L_a}\left[ \,
\bm{T}_{1 2}\sin (2\theta_2)
-\bm{T}_{1 3}\sin (2\theta_3)
\right]   \, ,
\\
\delta_{32}-\xi_{32}+z\xi'_{32}
=&\,
\frac{z^3}{2L_a}\left[\, 
\bm{T}_{1 2}(\sin\theta_2)^2\cos\theta_2
-\bm{T}_{1 3}(\sin\theta_3)^2\cos\theta_3
\right] \, , 
\end{split}
\end{align}
with solutions 
\begin{align}
\begin{split}
\lambda_{32}=&\,
\frac{
i(-2+\omega^2 z^2 (\cos\theta_2)^2)
}{
2\omega^3 L_a \cos\theta_2
}\, \bm{T}_{1 2}
-
\frac{
i(-2+\omega^2 z^2 (\cos\theta_3)^2)
}{
2\omega^3 L_a \cos\theta_3
} \,\bm{T}_{1 3}
+ib_2 e^{-i\omega z}  \, ,\\
\delta_{32}=&\,
-\frac{
i(
2+ \omega^2 z^2 (\cos\theta_2)^2
-2i\omega z\sin\theta_2
)\tan\theta_2
}{
2\omega^3 L_a
} \,\bm{T}_{1 2}  \, \\
&+\frac{
i(
2+ \omega^2 z^2 (\cos\theta_3)^2
-2i\omega z\sin\theta_3
)\tan\theta_3
}{
2\omega^3 L_a
} \,\bm{T}_{1 3}
+b_2 (-i+\omega z) e^{-i\omega z} \, ,\\
\xi_{32}=&\,
-\frac{
i(
4i\omega z\cos (2\theta_2)
+(8+\omega^2 z^2)\sin\theta_2
+\omega z(4i+\omega z\sin (3\theta_2))
)
}{
8\omega^3 L_a \cos\theta_2
}\, \bm{T}_{1 2}  \, \\
&+
\frac{
i(
4i\omega z\cos (2\theta_3)
+(8+\omega^2 z^2)\sin\theta_3
+\omega z(4i+\omega z\sin (3\theta_3))
)
}{
8\omega^3 L_a \cos\theta_3
} \,\bm{T}_{1 3}
-ib_2 e^{-i\omega z}  \, ,
\end{split}
\end{align}
where $b_2$ is an integration constant and we have used the infalling-wave boundary condition at IR.
The sourceless condition 
$\lambda_{32}(0)=\delta_{32}(0)=0$
give
\begin{align}
\mathcal{T}_{1 3}=
\mathcal{T}_{1 2}
\frac{\cos\theta_3}{1+\sin\theta_3}
\left(
\frac{\cos\theta_2}{1+\sin\theta_2}
\right)^{-1} \, ,\ \ 
b_2=-
\frac{
\sin\theta_2-\sin\theta_3
}{
\omega^3 L_a (1+\sin\theta_3) \cos\theta_2
}
\mathcal{T}_{1 2} \, ,
\end{align}
and $\xi_{32}(0)=0$ can be derived from $\lambda_{32}(0)=\delta_{32}(0)=0$.

The first order of equation \eqref{eq:q} gives only one independent equation
\begin{align}
(1+\cot\theta_1 \tan\theta_2+\cot\theta_3 \tan\theta_2)
\delta_2
+
(
\cot\theta_1 \ \delta_{12}
+\cot\theta_3 \ \delta_{32}
)=
\frac{i}{4L_a\omega^3}\,
\bm{S} \, , \label{eq:1st-7}
\end{align}
where
\begin{align}
\begin{split}
\bm{S}=&-
(
4+\omega^2 z^2+\omega^2 z^2\cos(2\theta_1)
+4i\omega z\sin\theta_1
)\,
(\bm{I}-\bm{R}_1)  \\
&+
(
4+\omega^2 z^2+\omega^2 z^2\cos(2\theta_2)
-4i\omega z\sin\theta_2
)\,\bm{T}_{1 2}  \\
&+
(
4+\omega^2 z^2+\omega^2 z^2\cos(2\theta_3)
-4i\omega z\sin\theta_3
)\,\bm{T}_{1 3}  \, .
\end{split}
\end{align}
The other two equations from \eqref{eq:q} are independent of \eqref{eq:1st-2}, \eqref{eq:isreal2} and \eqref{eq:1st-7}. 
Using the boundary conditions $\delta_2(0)=\delta_{12}(0)=\delta_{32}(0)=0$, the  $z\to 0$ limit of \eqref{eq:1st-7} yields the energy conservation equation
\begin{align}
\mathcal{T}_{1 2}+\mathcal{T}_{1 3}
+\mathcal{R}_1=1  \, .
\end{align}
Substituting the solutions for $\delta_{12}$ and  $\delta_{32}$ in \eqref{eq:1st-7}, we obtain
\begin{align}
\delta_2=
\frac{
i(
2+\omega^2 z^2 (\cos\theta_2)^2-2i\omega z\sin\theta_2
)
}{
2\omega^3 L_a
}\, \bm{T}_{1 2}  
+
\frac{
\left[ 
b_1 (1+i\omega z)\cot\theta_1
+b_2 (i-\omega z)\cot\theta_3
\right] 
e^{-i\omega z}
}{
1+(\cot\theta_1+\cot\theta_3)\tan\theta_2
}  \, .
\end{align}
The sourceless condition $\delta_2 (0)=0$ gives
\begin{align}
\frac{
i
}{
\omega^3 L_a
}  \mathcal{T}_{1 2}
+
\frac{
b_1 \cot\theta_1
+i b_2 \cot\theta_3
}{
1+(\cot\theta_1+\cot\theta_3)\tan\theta_2
}=0\,.
\end{align}

The infalling boundary condition at IR and  sourceless condition determine the three transport coefficients
\begin{align}
\mathcal{T}_{12}=
\frac{2s_2 }{(1+\sin\theta_1) C_s}  \,  ,~~~\ 
\mathcal{T}_{13}=
\frac{2s_3 }{(1+\sin\theta_1) C_s}  \,  ,~~~\ 
\mathcal{R}_1=1-(\mathcal{T}_{12}
+\mathcal{T}_{13})  \, , \label{eq:trans_3CFT}
\end{align}
where 
\begin{align}
\label{eq:sacs}
s_\text{A}=\frac{\cos\theta_\text{A}}{1+\sin\theta_\text{A}} \, ,\ \ ~~~
C_s=
\sum_{\text{A}=1}^3 s_\text{A}  \,  .
\end{align}
The coefficient $\mathcal{T}_{1 3}$ can alternatively be obtained by exchanging $\theta_2$ and $\theta_3$ in the expression for  $\mathcal{T}_{1 2}$. Similarly, all the other transport follow through index permutations   
\bea
\begin{split}\label{eq:trans_3CFT2}
&\mathcal{T}_{21}=
\frac{2s_1 }{(1+\sin\theta_2) C_s}  \,  ,~~~\ 
\mathcal{T}_{23}=
\frac{2s_3 }{(1+\sin\theta_2) C_s}  \,  ,~~~\ 
\mathcal{R}_2=1-(\mathcal{T}_{21}
+\mathcal{T}_{23})  \, ,\\
&\mathcal{T}_{31}=
\frac{2s_1 }{(1+\sin\theta_3) C_s}  \,  ,~~~\ 
\mathcal{T}_{32}=
\frac{2s_2 }{(1+\sin\theta_3) C_s}  \,  ,~~~\ 
\mathcal{R}_3=1-(\mathcal{T}_{31}
+\mathcal{T}_{32})  \, .
\end{split}
\eea
Equations 
\eqref{eq:trans_3CFT} and \eqref{eq:trans_3CFT2} provide complete expressions for energy transport. 
Notably, they automatically satisfy the energy conservation equation 
\be
\mathcal{R}_\text{A}+\sum_{\text{B}\neq \text{A}}\mathcal{T}_{\text{A}\text{B}}=1\,.
\ee
It is easy to verify that the holographic results presented above satisfy  \eqref{eq:rel1} and \eqref{eq:rel2} in Sec. \ref{sec:ft}. 

We first verify the consistency of the above formulae \eqref{eq:trans_3CFT} and \eqref{eq:trans_3CFT2} by considering the decoupling limit 
\begin{align}
\theta_3\to\frac{\pi}{2} \, ,\ \ 
L_3\to 0 \, , \ \ 
L_a \ \text{is finite} \, ,\label{eq:icft_limit}
\end{align}
where CFT$_3$ decouples from the other two CFTs. In this limit, the transport coefficients simplify to
\begin{align}
\begin{split}
\mathcal{T}_{1 2}=
\frac{
2\cos\theta_2
}{
\cos\theta_2 (1+\sin\theta_1)+
\cos\theta_1 (1+\sin\theta_2)
} \, , \  ~~
\mathcal{T}_{1 3}=0 \, , \  ~~
\mathcal{R}_1= 
1-\mathcal{T}_{1 2} \,  , \\ 
\mathcal{T}_{2 1}= 
\frac{
2\cos\theta_1
}{
\cos\theta_2 (1+\sin\theta_1)+
\cos\theta_1 (1+\sin\theta_2)
} \, , \  ~~
\mathcal{T}_{2 3}=0 \, , \ ~~
\mathcal{R}_2= 
1-\mathcal{T}_{2 1}  \, .
\end{split}
\end{align}
The results of $\mathcal{T}_{1 2},\mathcal{T}_{21},\mathcal{R}_1,\mathcal{R}_2$ exactly match the known 
ICFT$_2$ transmission coefficient  \cite{Bachas:2020yxv}, with the vanishing of $\mathcal{T}_{1 3}$ and $\mathcal{T}_{2 3}$ being consistent with $L_3\to 0$. Note that 
in the limit \eqref{eq:icft_limit}, the transport coefficients $\mathcal{T}_{3 1},\mathcal{T}_{32},\mathcal{R}_3$ are meaningless since the amplitude of incoming wave in CFT$_3$ is proportional to $L_3$.

Another interesting limit is to consider the large tension limit. Following \cite{Anous:2022wqh}, we expand the configuration near maximal tension,  
\be 
\label{eq:largetension}
T=T_\text{max}-\frac{1}{2(L_1+L_2+L_3)}\delta^2  \, ,
\ee
the angular expansion are determined from the junction condition 
\be
\label{eq:largetension2}
\theta_A=\frac{\pi}{2}-\frac{L_A}{L_1+L_2+L_3}\delta
+\mathcal{O}(\delta^2)
\,,
\ee
with $\delta\to 0$. 
To the leading order in $\delta$, the transport coefficients become
\bea
\label{eq:translargeT}
\begin{split}
\mathcal{T}_{1 2}=\mathcal{T}_{32}=\frac{L_2}{L_1+L_2+L_3}
+\mathcal{O}(\delta^2)
=\mathcal{R}_{2}\,,\\
\mathcal{T}_{13}=\mathcal{T}_{23}=\frac{L_3}{L_1+L_2+L_3}
+\mathcal{O}(\delta^2)=\mathcal{R}_{3}\,,\\
\mathcal{T}_{21}=
\mathcal{T}_{31}=\frac{L_1}{L_1+L_2+L_3}
+\mathcal{O}(\delta^2)=\mathcal{R}_{1}\, .
\end{split}
\eea  
As we will discuss in detail in Sec. \ref{sec:etindetail}, these values are the extremal bounds for the transmission coefficients.

\subsubsection{Energy transport bounds from quantum information quantities}
\label{subsec:bound}

For conformal interface, it is known that energy transmission is bounded by information transmission \cite{Karch:2024udk}. Specifically, the inequality 
$0\leq c_{LR}\leq c_\text{eff} \leq \text{min}(c_L, c_R)$ holds for conformal interface, where $ c_{LR}$ characterizes the energy transmission while $c_\text{eff}$ and $c_L, c_R$ are quantum information measure. 

Following the discussion in Sec. \ref{sec:ft}, we define the coefficients
\begin{align}
c_\text{AB}\equiv c_\text{A} \mathcal{T}_{\text{A}\text{B}}
=\frac{2c_a}{C_s}s_\text{A} s_\text{B}\, ,
\end{align}
where $c_a$ is given in \eqref{eq:ca}, and $C_s$,  $s_A$ are defined in \eqref{eq:sacs}. Obviously, these coefficients satisfy
\begin{align}
c_\text{AB}=c_\text{BA}  \,  ,
\end{align}
which corresponds to the detailed-balance condition discussed in \cite{Bachas:2021tnp} as well as the discussion of \eqref{eq:rel3} in Sec. \ref{sec:ft}. From the angular region $-\frac{\pi}{2} <\theta_\text{A}<\frac{\pi}{2}$\,, we know 
\begin{align}
 c_\text{AB}>0 \, .
\end{align}

To establish an upper bound for the coefficient $c_{AB}$ using quantum information measure, we must partition the spatial region of the dual ICFT$_3$ into two subsystems $A$ and $BC$ (where $B, C\neq A$). This requires calculating the entropy for region $A$, which gives result of $c_A$ as well as the entropy contribution from $B$. This naturally reminds us the setup of multi-entropy developed in \cite{Gadde:2022cqi, Penington:2022dhr, Ju:2024hba, Ju:2024kuc, Basak:2024uwc}, where the three systems $A, B, C$ are considered which may correspond to  GHZ-state or W-state. It is expected that $c_{AB}$ is bounded by $c_A$ and the multiple entropy of region $B$, potentially yielding a tighter constraints than  \eqref{eq:conscab}.  We leave this investigation for future work. 

Here we focus on the total transmission coefficient, which is defined as  
\be
\mathcal{T}_A=\sum_{B\neq A}\mathcal{T}_{AB}=1-\mathcal{R}_A\,.
\ee

The effective central charge $c_\text{eff}$ characterizes the entanglement entropy between the $A$-th CFT and the remaining $(N-1)$ CFTs
\begin{align}
S^\text{A}_{E}=\frac{c^\text{A}_\text{eff}}{6}
\log\frac{\ell_\text{IR}}{\epsilon_\text{UV}} \, ,
\end{align}
where $\ell_\text{IR},\epsilon_\text{UV}$ are IR and UV cutoffs respectively, and 
\be
c^\text{A}_{\text{eff}}=\text{min}
\left(
c_A, \sum_{\text{B}\not= \text{A}} c_\text{B}
\right)\, .
\ee
In our holographic setup, this inequality is saturated, similar to the behavior observed in conformal interfaces \cite{Karch:2022vot, Karch:2023evr}. 
For general ICFT$_3$, we expect that this bound to be tight, i.e. $c^\text{A}_{\text{eff}}\leq \text{min}
\big(
c_A, \sum_{\text{B}\not= \text{A}} c_\text{B}
\big).$ 

Therefore, the energy transport is expected to be  constrained by the following inequalities 
\be\label{eq:bound}
c_A \mathcal{T}_A\leq c^\text{A}_\text{eff}\,,
\ee
i.e.
\be
c_1 \mathcal{T}_1 \leq \text{min}(c_1, c_2+c_3)\,,~~c_2 \mathcal{T}_2 \leq \text{min}(c_2, c_1+c_3)\,,~~
c_3 \mathcal{T}_3 \leq \text{min}(c_3, c_1+c_2)\,. 
\ee
We will verify that these bounds hold for the configurations under consideration.

\subsubsection{Comment on island formulae in ICFT$_3$}

There exist three distinct descriptions on ICFTs in the AdS/ICFT correspondence
: (1) the 3D bulk gravity with a junction  interface brane, (2) a 2D gravitational system on the interface brane coupled to three  non–gravitating bath CFTs, and (3) a purely 2D ICFT description   \cite{Anous:2022wqh}.  In the large tension limit, the second description is expected to be equivalent to the other,  {\em i.e.} the system can be viewed as a gravitational theory lived on the junction brane coupled to three separate CFTs defined on half-lines. From this perspective, we can study the entanglement entropy for spatial regions $[0, \sigma_A)$ in ICFT$_3$ using the island formula \cite{Almheiri:2020cfm}
\be
S_\text{island}=\text{min}_z \sum_\text{A}\frac{c_\text{A}}{6}\log\left(
\frac{(z+\sigma_\text{A})^2}{z\tilde{\epsilon} \cos\theta_\text{A}}
\right)\,.
\ee
The position of quantum extremal surface $z_*$ is determined by 
\be
\sum_A \cos\theta_\text{A}
\frac{z-\sigma_\text{A}}{z(z+\sigma_\text{A})}
=0  \, .
\ee
While no general analytical solution exists for arbitrary $\sigma_A$, the symmetric case  $\sigma_1=\sigma_2=\sigma_3=\sigma$ yields $z_*=\sigma$. 
Thus 
\be
\label{eq:island}
S_\text{island}=-\sum_A\frac{c_A}{6}
\log\delta+\sum_A\frac{c_A}{6}
\log\left[\frac{(z_*+\sigma_A)^2}{z_*\tilde{\epsilon}}\left(\frac{L_1+L_2+L_3}{L_A}\right)\right]+\mathcal{O}(\delta)\,,\ee
 where $\delta\to 0$ is a parameter that characterises the large tension limit introduced  in Eq.  \eqref{eq:largetension}. 

Then entanglement entropy for regions $[0, \sigma_A)$ has been independently calculated in appendix \ref{app:b} and takes the form of \eqref{eq:entropy}. In the large tension limit defined by \eqref{eq:largetension} and \eqref{eq:largetension2},  we find
\be\label{eq:entropylarget2}
S_E=
-\sum_A\frac{c_A}{6}
\log\delta+\sum_A\frac{c_A}{6}
\log\left[
\frac{4\sigma (L_1+L_2+L_3)}{L_\text{A}\delta\epsilon}
\left(
\frac{L_1+L_2+L_3}{L_A}
\right)
\right]+\mathcal{O}(\delta)\,.
\ee
By identifying the cutoff scale 
$\tilde{\epsilon}=
\epsilon\delta\prod_{\text{A}=1}^3
\left(
\frac{L_\text{A}}{L_1+L_2+L_3}
\right)
^{\frac{L_\text{A}}{L_1+L_2+L_3} }
$, 
we obtain exact agreement with the island formula result \eqref{eq:island}. This provides strong supporting evidences for the validity of the second description outlined at the beginning of this subsection in the context of the holographic junctions of ICFT$_3$.  


\subsection{Energy transport in different parameter regions}
\label{sec:etindetail}

In this section, we systematically examine energy transport within different parameter regimes of tension and bulk AdS radii. Following our earlier discussion in Sec. \ref{subsec:rangetension}, we specifically focus on the tension ranges \eqref{eq:tensionran1} and \eqref{eq:tensionran2}, analyzing each case separately.

\subsubsection{$L_1 \leq L_2<L_3$}

We first analyze the case where the AdS radii satisfy $L_1\leq L_2<L_3$ with  $L_1+L_2<L_3$, focusing on the tension range 
\be
\label{eq:tension1}
\frac{1}{L_1}+\frac{1}{L_2}-\frac{1}{L_3}<T<\frac{1}{L_1}+\frac{1}{L_2}+\frac{1}{L_3}  \, ,
\ee
which guarantees a unique bulk configuration for each tension value. Within this regime, we analytically find 
\be
\frac{d\log g}{dT}>0 \, ,
\ee
and the range of $\log g$ is
\be
\log g\in (-\infty, +\infty)  \, .
\ee
This feature is also observed in AdS/ICFT$_2$ \cite{Simidzija:2020ukv}. 

\begin{figure}[h!]
\begin{center}
\includegraphics[width=0.49\textwidth]{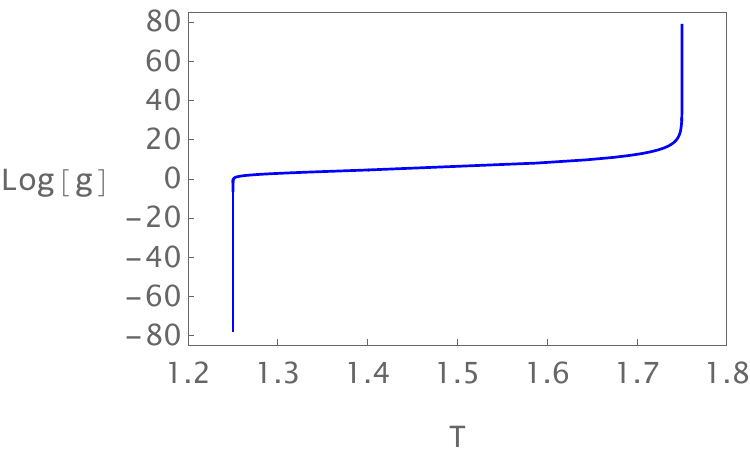}
~~~~~
\includegraphics[width=0.43\textwidth]{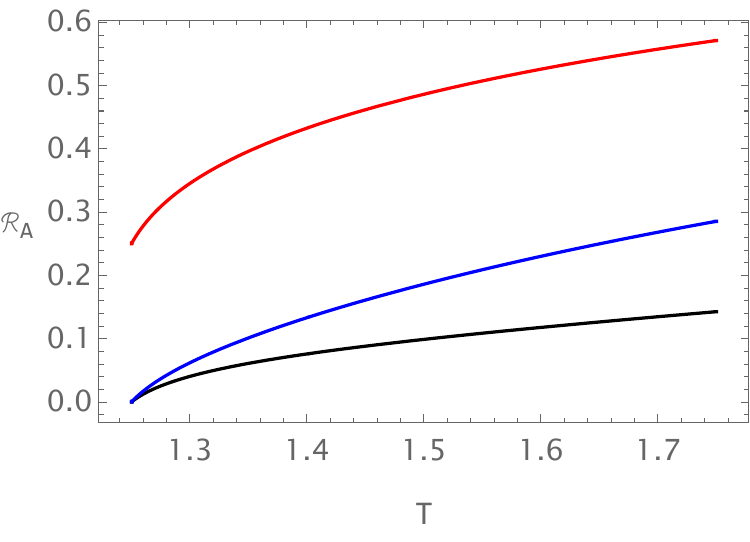}
\end{center}
\vspace{-0.3cm}
\caption{\small  Plots of junction entropy  $\log g$ ({\em left}) and energy  reflection coefficients ({\em right}) including $\mathcal{R}_1$ ({\em black}), $\mathcal{R}_2$ ({\em blue}), $\mathcal{R}_3$ ({\em red}), as functions  of tension $T$ for $L_1=1,L_2=2,L_3=4$.
}
\label{fig:reflection1}
\end{figure}

Before detailed analysis, we first show a  numerical result on the $\log g$ and the energy reflection coefficient for ICFT$_3$ as a function of the tension. The right plot in Fig. \ref{fig:reflection1} shows that all the energy reflection coefficients are monotonic increasing function of the tension.

We know that near the upper bound of the tension in \eqref{eq:tension1}, the transport coefficients are given by \eqref{eq:translargeT}. Near the lower bound of the tension in \eqref{eq:tension1}, we consider the limit 
\be
T=\frac{1}{L_1}+\frac{1}{L_2}
-\frac{1}{L_3}-\frac{\delta^2}{2(L_1+L_2-L_3)}
\ee 
with angular parameters
\bea
\begin{split}
\theta_1&=\frac{\pi}{2}+\frac{L_1}{L_1+L_2-L_3}\delta+\mathcal{O}(\delta^2)\,,\\
\theta_2&=\frac{\pi}{2}+\frac{L_2}{L_1+L_2-L_3}\delta+\mathcal{O}(\delta^2)\,,\\
\theta_3&=-\frac{\pi}{2}-\frac{L_3}{L_1+L_2-L_3}\delta+\mathcal{O}(\delta^2)\,,
\end{split}
\eea
which satisfy the junction condition \eqref{eq:bg jc}.  
The resulting transport coefficients are 
\bea
\begin{split}
&\mathcal{T}_{1 2}=\frac{L_2L_3}{4(L_1+L_2-L_3)^2}\delta^2\,,~~~~~~~~~
\mathcal{T}_{32}=\frac{L_2}{L_3}+\frac{L_2(L_2^2-
3 (L_1+L_2) L_3+2 L^2_3
)}{12L_3(L_1+L_2-L_3)^2}\delta^2\,,\\
&\mathcal{T}_{13}=1+\frac{L_1^2-(L_1+L_2)L_3}{4(L_1+L_2-L_3)^2}\delta^2\,,~~~~\mathcal{T}_{23}=1+\frac{L_2^2-(L_1+L_2)L_3}{4(L_1+L_2-L_3)^2}\delta^2\,,~~~\\
&\mathcal{T}_{21}=\frac{L_1L_3}{4(L_1+L_2-L_3)^2}\delta^2\,,~~~~~~~~~\mathcal{T}_{31}=\frac{L_1}{L_3}+\frac{L_1(L_1^2-
3 (L_1+L_2) L_3 +2 L_3^2
)}{12L_3(L_1+L_2-L_3)^2}\delta^2\,,\\
&\mathcal{R}_{1}=\frac{L_1(L_3-L_1)}{4(L_1+L_2-L_3)^2}\delta^2\,,~~~~~~~~~\mathcal{R}_{2}=\frac{L_2(L_3-L_2)}{4(L_1+L_2-L_3)^2}\delta^2\,,\\
&\mathcal{R}_{3}=\frac{L_3-L_1-L_2}{L_3}-
\frac{
(L_1+L_2)
(
L_1^2+L_2^2-3 (L_1+L_2) L_3 +2 L_3^2 -L_1 L_2
)
}{
12(L_1+L_2-L_3)^2 L_3
} \delta^2
\, .
\end{split}
\eea 

We have checked that all transmission coefficients vary  monotonically within the tension range  \eqref{eq:tensionran1}. Specially, as tension increases, $\mathcal{T}_{12}$ and $\mathcal{T}_{21}$ increase monotonically, while  $\mathcal{T}_{13}, \mathcal{T}_{31}, \mathcal{T}_{23}$ and $\mathcal{T}_{32}$ decrease monotonically. The allowed ranges for energy reflection coefficients are 
\bea
\mathcal{R}_{1}\in \left(0,~\frac{L_1}{L_1+L_2+L_3}\right)\,,
~\mathcal{R}_{2}\in 
\left(0,~\frac{L_3}{L_1+L_2+L_3}\right)\,,
~\mathcal{R}_{3}\in \left(\frac{L_3-L_1-L_2}{L_3},~1\right)\,, 
\nn
\eea
or equivalently for energy transmission coefficients 
\bea
\mathcal{T}_{1}\in \left(\frac{L_2+L_3}{L_1+L_2+L_3},~1\right)\,,
~~~\mathcal{T}_{2}\in 
\left(\frac{L_1+L_2}{L_1+L_2+L_3},~1\right)\,,
~~~\mathcal{T}_{3}\in \left(0,~\frac{L_1+L_2}{L_3}\right)\,.
\eea

Here we summarize our findings, 
\begin{itemize}
\item  For the tension range \eqref{eq:tension1}, the bulk solution is uniquely determined, with junction entropy $\log g$ being a monotonically increasing function of  tension that spans all real numbers. 
\item The energy reflection coefficients $\mathcal{R}_{1}, \mathcal{R}_{2}, \mathcal{R}_{3}$ (or transmission coefficients $\mathcal{T}_{1}, \mathcal{T}_{2}, \mathcal{T}_{3}$) are monotonic decreasing (increasing) functions of tension $T$ within \eqref{eq:tension1}.
\item The inequality  $c_\text{A}\mathcal{T}_\text{A}\le c^\text{A}_\text{eff}$ always holds. 
\end{itemize}

\subsubsection{$L_1\leq L_2=L_3$}

We first consider $L_1< L_2=L_3$. In this case we consider the tension range 
\be
\label{eq:tensionran3}
\frac{1}{L_1} \leq T<\frac{1}{L_1}+\frac{2}{L_0}  \, ,
\ee
where we have defined $L_0\equiv L_2=L_3$. 
The solution of the system is 
\begin{align}
\theta_1>0\,,~~~
\theta_2=\theta_3=\theta>0 \,, 
\end{align} satisfying 
\be
T=\frac{\sin\theta_1}{L_1}+\frac{2\sin\theta}{L_0}\,,~~~L_a=\frac{L_1}{\cos\theta_1}=\frac{L_0}{\cos\theta}\,.
\ee
The allowed range for $L_a$ is 
\be
L_a\in \left[ L_a^\text{min},\, +\infty \right) \, ,
\ee
where the minimal value is 
\be
L_a^\text{min}=
\sqrt{
\frac{
L^2_0 (2 L^2_0-3 L^2_1)+L^3_0
\sqrt{
4L^2_0-3 L^2_1
}
}{
4(L^2_0-L^2_1)
}
}  \, .
\ee
The case $L_1=L_0$ will be discussed later.

The $g$-function can be calculated to be 
\be
\log g= L_1 \log
\frac{
L_a+\sqrt{L^2_a-L^2_1}
}{
L_1
}  +
2 L_0 \log
\frac{
L_a+\sqrt{L^2_a-L^2_0}
}{
L_0
}  \, ,
\ee
which satisfies that $\frac{d\log g}{dL_a}>0$, leading to the range, 
\be\label{eq:loggcase2}
\log g\in \left[ 
L_1 \log
\frac{
L_a^\text{min}+\sqrt{(L_a^\text{min})^2-L^2_1}
}{
L_1
}  +
2 L_0 \log
\frac{
L_a^\text{min}+\sqrt{(L_a^\text{min})^2-L^2_0}
}{
L_0
}
\,,~~ +\infty\right) \, .
\ee
 
\begin{figure}[h!]
\begin{center}
\includegraphics[width=0.49\textwidth]{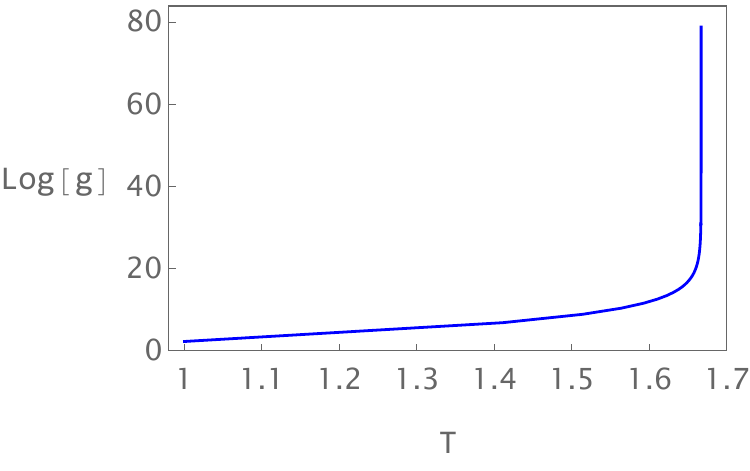}
~~~~~
\includegraphics[width=0.44\textwidth]{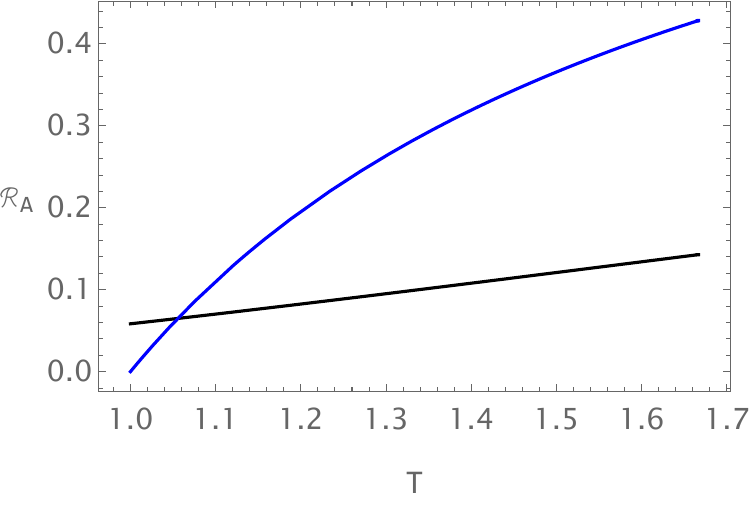}

\end{center}
\vspace{-0.3cm}
\caption{\small  
Plots of junction entropy $\log g$ ({\em left}) and energy reflection coefficients  ({\em right}), including $\mathcal{R}_1$ ({\em black}) and  $\mathcal{R}_2=\mathcal{R}_3$ ({\em blue}), as functions of tension $T$ for  $L_1=1,L_2=L_3=3$. 
}
\label{fig:reflection2}
\end{figure}

In Fig. \ref{fig:reflection2} we show the dependence of both the junction energy $\log g$ and the energy reflection coefficients on the tension.
These quantities again exhibit monotonic growth with increasing tension.  Unlike  from the previous case, $\log g$ here 
has a finite minimal value. 

Analytical computation reveals the following monotonic behaviors for the tension range \eqref{eq:tensionran2}, 
\begin{align}
\frac{d\mathcal{R}_\text{A}}{dT}>0 \, ,\ \ 
\frac{d\mathcal{T}_\text{AB}}{dT}<0 \, ,\ \ 
\text{A}=1,2,3 \ \ \text{and} \ \ \text{B}\not=\text{A}
\, .
\end{align}
This allows us to determine the ranges of all transport coefficients,
\bea
\begin{split}
&\mathcal{T}_{12}=\mathcal{T}_{13}\in \left(
\frac{L_0}{2L_0+L_1} \, ,~~
\mathcal{T}_{12}^\text{max}
\right]  \, , \\
&\mathcal{T}_{21}=\mathcal{T}_{31}\in\left(
\frac{L_1}{2L_0+L_1} \, ,~~
\mathcal{T}_{21}^\text{max}
\right ] \, ,\\
&\mathcal{T}_{23}=\mathcal{T}_{32}\in\left(
\frac{L_0}{2L_0+L_1} \, ,~~\mathcal{T}_{23}^\text{max}
\right] \, ,\\
&\mathcal{R}_{1}\in \left[\mathcal{R}_1^\text{min}\,,~~\frac{L_1}{2L_0+L_1}\right)
\, ,
\\
&\mathcal{R}_{2}=\mathcal{R}_{3}\in \left[ 0\,,~~\frac{L_0}{2L_0+L_1}\right) \, ,
\end{split}
\eea
where the maximal/minimum values are given by 
\begin{align}
\begin{split}
&\mathcal{T}_{12}^\text{max}=
\frac{
2L_0
}{
L_1 (1+\sqrt{1+4L^2_0 \gamma})
+2L_0 (1+\sqrt{1-4L^2_1\gamma})
}  \, ,\\
&\mathcal{T}_{21}^\text{max}=
\frac{
2L_1
}{
L_1 (1+\sqrt{1+4L^2_0 \gamma})
+2L_0 (1+\sqrt{1-4L^2_1\gamma})
}  \, ,\\
&\mathcal{T}_{23}^\text{max}=
1-\mathcal{T}_{21}^\text{max}  \, ,\\
&\mathcal{R}_1^\text{min}=
1-2 \mathcal{T}_{12}^\text{max} \,,\\
&\gamma\equiv
\frac{
L^2_1-L^2_0
}{
L^2_0 (2 L^2_0-3L^2_1)+
L^3_0 \sqrt{4L^2_0-3L^2_1}
}  \,  .
\end{split}
\end{align}

Equivalently, we obtain 
\be
\mathcal{T}_{1}\in \left(\frac{L_0-L_1}{2L_0+L_1}\,,~~
2\mathcal{T}_{12}^\text{max}\right] \,,~~
\mathcal{T}_{2}=\mathcal{T}_{3}\in \left(\frac{L_0-L_1}{2L_0+L_1}\,,~1\right] \,.
\ee
The relation \eqref{eq:bound} is clearly  satisfied, i.e. 
$
c_1\mathcal{T}_1 < c^1_\text{eff} \, ,~ 
c_2\mathcal{T}_2 \le c^2_\text{eff} \, ,~ 
c_3\mathcal{T}_3 \le c^3_\text{eff}  \, .
$

When $L_1=L_2=L_3$, we have $\theta_1=\theta$ and $L_a\in \big[\frac{3\sqrt{2}}{4}L_0\,,~ \infty\big)$. 
In the tension regimes 
\be 1\le TL_1<3\,, \label{eq:L1=L2=L3} \ee 
the configuration of the system is unique for each value of $TL_1$. The junction entropy satisfies 
\be
\label{eq:gran3}
\log g\in \left[~\frac{3}{2}L_1\log 2,~+\infty\right)\,.
\ee
We have exacted analytical expression for all the energy transport coefficients 
\begin{align}
\mathcal{T}_\text{AB}
=\frac{2}{TL_1+3} ~(\text{for~} \text{A}\neq \text{B})\, , \ \
~~\mathcal{T}_\text{A}=\frac{4}{TL_1+3}\,,~~~
\mathcal{R}_\text{A}
=\frac{TL_1-1}{TL_1+3}  \, ,
\end{align}
with corresponding ranges
\begin{align} 
\frac{2}{3}<
\mathcal{T}_\text{A}\le  1 \, ,\ \
0\le 
\mathcal{R}_\text{A} < \frac{1}{3}  \, .
\end{align}
The inequality \eqref{eq:bound} is  
satisfied.

We summarize our findings for $L_1\leq L_2=L_3\,$, 
\begin{itemize}
\item  Within tension regime \eqref{eq:tensionran2} (and \eqref{eq:L1=L2=L3}), the bulk solution is unique and $\log g$ increases monotonically with the tension, spanning the range \eqref{eq:loggcase2} (and \eqref{eq:gran3}). 
\item
    All energy reflection coefficients  $\mathcal{R}_\text{A}$ and energy transmission coefficients $ \mathcal{T}_\text{AB}$ vary monotonically with $T$ (increasing and decreasing respectively). 
\item  The bound  $c_\text{A}\mathcal{T}_\text{A} \le  c^\text{A}_\text{eff}$ holds universally.
\end{itemize}

\section{Generalization to $N$ CFT junctions}
\label{sec:n}
We now extend our analysis to holographic  conformal junctions connecting $N\geq 3$ CFTs. The generalization of ICFT$_3$ to ICFT$_N$ follows straightforwardly, and we present  the key results below.  

The action and background ansatz generalize directly from \eqref{eq:action} and \eqref{eq:ads3metric}, replacing $3$ with $N$. The holographic background solution satisfies 
\begin{align}
\frac{L_\text{1}}{\cos\theta_{1}}=
\frac{L_\text{B}}{\cos\theta_\text{B}} \, ,~~~\ \ 
\text{B}=2, 3,\cdots, N \, ,
\end{align}
and
\begin{align}
T=\sum_{\text{A}=1}^N \frac{\sin\theta_\text{A}}{L_\text{A}}  \,  .
\end{align}

Following the same procedure for the derivation of \eqref{eq:R1T2} in AdS/ICFT$_3$, the first order continuity  condition
\be
g_{ab}^\text{A}\frac{\partial x_\text{A}^a}{\partial y^\mu}
\frac{\partial x_\text{A}^b}{\partial y^\nu}\bigg{|}_\text{A=1}
=\, g_{ab}^\text{A}\frac{\partial x_\text{A}^a}{\partial y^\mu}
\frac{\partial x_\text{A}^b}{\partial y^\nu}\bigg{|}_\text{A=B}  \, ,\ \ ~~
\text{B}=2, 3,\cdots, N \, ,
\ee
gives
\begin{align}
\mathcal{R}_1=&
\frac{
-1+\sin\theta_1
}{
1+\sin\theta_1
}
+\mathcal{T}_{1 \text{B}}
\frac{\cos\theta_1}{1+\sin\theta_1}
\left(
\frac{\cos\theta_\text{B}}{1+\sin\theta_\text{B}}
\right)^{-1}  \, ,\ \ ~~
\text{B}=2, 3,\cdots, N \, . \label{eq:R1TBforn}
\end{align}

The first order junction condition 
$
\sum_{\text{A}=1}^N K^{\text{A}}=2T  \, 
$
simplifies to 
\begin{align}
\omega^2 z \delta_t -2\delta'_t +z\delta''_t
=\frac{1}{2L_a}\bm{S}_N \, , \label{eq:delt_t}
\end{align}
where
\begin{align}
\delta_t=
\sum_{\text{A}=1}^N \delta_\text{A} \, ,\ \ 
\bm{S}_N=
-i\omega z^3 (\cos\theta_1)^4 \,
(\bm{I}-\bm{R}_{1}) 
+i\omega z^3
\sum_{\text{B}=2}^N
(\cos\theta_\text{B})^4\, \bm{T}_{1 \text{B}} 
\, .
\end{align}
The solution to \eqref{eq:delt_t} is
\begin{align}
\begin{split}
\delta_t=&
-i\frac{
2+\omega^2 z^2 (\cos\theta_1)^2 
+2i\omega z\sin\theta_1
}{
2\omega^3 L_a
}\,\bm{I}
+
i\frac{
2+\omega^2 z^2 (\cos\theta_1)^2 
-2i\omega z\sin\theta_1
}{
2\omega^3 L_a
}\, \bm{R}_1  \, \\
&+i
\sum_{\text{B}=2}^N
\frac{
2+\omega^2 z^2 (\cos\theta_\text{B})^2
-2i\omega z\sin\theta_\text{B}
}{
2\omega^3 L_a
} \,\bm{T}_{1 \text{B}}  \, ,
\end{split}
\end{align}
where we have droped the homogeneous solution $e^{\pm i\omega z}$ because $\delta_t$ is the perturbation in the normal direction of the brane and it should vanish in the absence of gravity waves $\bm{I},\bm{R}_1,\bm{T}_{1 \text{B}}$.
The sourceless condition
$
\delta_t (0)=0
$
gives the energy conservation equation
\begin{align}
\mathcal{R}_1+
\sum_{\text{B}=2}^N \mathcal{T}_{1 \text{B}}
=1  \, .\label{eq:ecn}
\end{align}

From \eqref{eq:R1TBforn} and \eqref{eq:ecn}, we 
obtain all energy transmission coefficients,
\begin{align}
\mathcal{T}_{\text{A} \text{B}}
=&\,\frac{2s_\text{B}}{(1+\sin\theta_\text{A})C_s}
 \,  ,~~~~~ (\text{B}\in \{ 1,2,\cdots,N\}\,\  \text{and}\  \text{B}\not=\text{A})   \label{eq:transN}  \, ,\\
\mathcal{R}_\text{A}=&\,
1-\sum_{\text{B}\not=\text{A}} \mathcal{T}_{\text{A}
\text{B}}
=\frac{1}{1+\sin\theta_\text{A}}\left[
-1+\sin\theta_\text{A}+\frac{2s_\text{A}}{C_s}
\right]
\, ,
\end{align}
where
\begin{align}
C_s=\sum_{\text{A}=1}^{N} s_\text{A}  \,,~~~~~ s_\text{A}=\frac{\cos\theta_\text{A}}{1+\sin\theta_\text{A}} .
\end{align}
For $-\frac{\pi}{2}<\theta_\text{A}<\frac{\pi}{2}$, we have
 $s_\text{A}>0,1+\sin\theta_\text{A}>0$ and consequently  
 $\mathcal{T}_{\text{A} \text{B}}>0$.
Here, $s_\text{B}$ acts as the weight for transmission components,
\begin{align}
\mathcal{T}_{\text{A} \text{B}}=
\frac{s_\text{B}}
{\sum_{\text{C}\not=\text{A}} s_\text{C}}
(1-\mathcal{R}_\text{A}) \, ,~~~~
(\text{B}\not=\text{A})
\, . \label{eq:refN}
\end{align}
From \eqref{eq:transN} and \eqref{eq:refN}, we find,
\begin{align}
\mathcal{T}_{\text{A} \text{B}}>0 \, ,\ \ 
\mathcal{R}_\text{A}<1 \, .
\end{align}

Following \cite{Meineri:2019ycm}, we  define the transmission coefficients 
\begin{align}
c_\text{AB}\equiv c_\text{A} \mathcal{T}_{\text{A} \text{B}}
=\frac{2c_a}{C_s}s_\text{A} s_\text{B}\, ,
\end{align}
where $c_a$ is given in \eqref{eq:ca}. These coefficients satisfy the symmetry relation
\begin{align}
c_\text{AB}=c_\text{BA}  \,  ,
\label{eq:Ndbc}
\end{align}
which is the same as the detailed-balance condition discussed in \cite{Bachas:2021tnp}.

The system has $N^2$ transport coefficients in total. 
The energy conservation equation \eqref{eq:ecn} gives $N$ constraints and the detailed-balance condition \eqref{eq:Ndbc} gives $N-1$ constrains, thus there are only $(N-1)^2$ independent transport coefficients. This matches exactly with the R-matrix analysis in Sec. \ref{sec:ft}.

We can further define a total transmission coefficient and establish its bounds using the effective central charge. These bounds can be explicitly verified using detailed  solutions. Although a detailed analyze of the bulk solution, the g-function behavior, and energy transport is straightforward and would be valuable, we  leave such detailed investigations for future work. 
\section{Conclusions and open questions}
\label{sec:cd}

We have investigated energy transport in holographic junctions connecting $N$ CFTs, extending previous studies of conformal interfaces ($N=2$). Focusing on AdS/ICFT$_3$, we found that within the tension range specified by \eqref{eq:tensionran1} (or \eqref{eq:tensionran2}) and the AdS radii satisfying relation $L_1\leq L_2<L_3$ and $L_1+L_2<L_3$ (or $L_1\leq L_2=L_3$), the bulk solution is uniquely determined. The junction entropy $\log g$ increases monotonically with tension and, remarkably, spans the full range $(-\infty, \infty)$ in tension regime \eqref{eq:tensionran1}, analogous to the conformal interface case  \cite{Simidzija:2020ukv}.  Through holographic calculations, we derived the energy transport coefficients for these cases, all of which are positive, bounded and vary monotonically with tension. Our analysis reveals a new inequality:
\be 
\label{eq:inequality2}
0\leq \sum_{B\neq A} c_{AB}\leq c^A_\text{eff}\leq \text{min}\Big(c_A,\, \sum_{B\neq A}c_B\Big)\,
\ee 
which generalizes the known inequality for conformal interface \cite{Karch:2024udk}. We further extend our study to holographic ICFT$_N$ 
and derived complete expressions for all energy transport coefficients.

While our evidences for this inequality \eqref{eq:inequality2} is derived from a specific holographic model, it would be valuable to verify it in other contexts, including field-theoretic  examples, such as the case of massless fermions \cite{Bellazzini:2008fu, Mintchev:2011mx},  generalized Janus solutions \cite{Bak:2007jm} for thick wall models and top-down holographic models for $N$ CFTs junctions. Additionally, extending the finite-temperature methods  for conformal interfaces \cite{Bachas:2021tnp} to study transport properties in holographic junctions would be worthwhile. 

There are several important directions for future research. 
It would be valuable to explore the constraints on $c_{AB}$ coefficients, which, may relate to multi-entropy measures. The role of multi-entropy in 
transport bound is likely analogous to the 
universal relation observed in conformal interfaces \cite{Karch:2022vot}.Moreover, based on the discussion in Sec. \ref{subsec:bound}
, one could define more different types of $c_\text{eff}$ for an $N$ junction system using  multi-entropy. It would be interesting to explore their potential roles in a generalized version of the inequality  \eqref{eq:inequality2}.  
It would also be interesting to understand better energy transport in the unexplored tension regimes, i.e. the gray curves in Fig.  \ref{fig:tension}, including investigating the stability of these configurations \cite{Czech:2016nxc}. 

Further studies on current transport in AdS/ICFT by probing a vector field in the bulk could have implications for condensed matter physics and conformal field theory.  While field theory studies of current transport in multiple junction systems exist \cite{Kimura:2015nka}, a holographic perspective may reveal new insights. Unlike energy transport, which is constrained by ANEC, current reflection or transmission coefficients may exhibit negative values. Finally, generalizing junction systems to  higher dimensional\footnote{ After this work appeared on arXiv, we  became aware of the study in  \cite{Guo:2025sbm}, which constructs the gravity dual of CFTs on a network 
in general dimensions. 
The multijunction along Euclidean time was studied in \cite{Jiang:2025iet}. 
} or non-conformal settings \cite{Liu:2024oxg} could provide new theoretical insights and potential experimental connections.

\vspace{.3cm}
\subsection*{Acknowledgments}
We thank Bartek Czech, Dong-Sheng Ge, Policastro Giuseppe, Andrew Karch, Cheng Peng and Ya-Wen Sun 
for useful discussions. This work is supported by the National Natural Science Foundation of China grant No. 12375041 and 12575046. 

\vspace{.6 cm}
\appendix
\section{Energy transport in holographic ICFT$_2$}
\label{app:a}
Here we review the results on energy transport in holographic ICFT$_2$ from \cite{Bachas:2020yxv}. 
For holographic interfaces, i.e. $N=2$, the background solution satisfies
\begin{align}
L_a=&
\frac{L_\text{1}}{\cos\theta_{1}}=
\frac{L_\text{2}}{\cos\theta_{2}} \,  ,~~~
T=\sum_{\text{A}=1}^2 \frac{\sin\theta_\text{A}}{L_\text{A}}  \, ,
\end{align}
with energy transmission coefficients  
\begin{align}
\mathcal{T}_{1 2}=\frac{2}{L_2}
\left[
\frac{1}{L_1}+\frac{1}{L_2}+T
\right]^{-1}   \, ,\ \ 
\mathcal{T}_{2 1}=\frac{2}{L_1}
\left[
\frac{1}{L_1}+\frac{1}{L_2}+T
\right]^{-1} \,, 
\end{align}
and energy reflection coefficients
\be
\mathcal{R}_1=1-\mathcal{T}_{1 2}\,,~~
\mathcal{R}_2=1-\mathcal{T}_{21}\,.~~
\ee

As discussed in Sec. \ref{subsec:rangetension},  negative tension is excluded by the null energy condition. We will focus on case of positive tension.

For $L_1<L_2$, the tension range is 
\begin{align}
T\in
\left(
\frac{1}{L_1}-\frac{1}{L_2} \,,~~
\frac{1}{L_1}+\frac{1}{L_2}
\right)
\, .
\end{align}
The interface entropy is 
\begin{align}
\log g=&\frac{1}{4G}\sum_{\text{A}=1}^{2}
\rho_\text{A}^{*} \, ,
\end{align}
where \begin{align}
\sin\theta_\text{A}=\tanh\frac{\rho_\text{A}^{*}}{L_\text{A}}  \, .
\end{align}
The range of the interface entropy is
\begin{align}
-\infty
<\log g<&+\infty
\, ,
\end{align}
and the corresponding range of transmission coefficients are 
\begin{align}
\mathcal{T}_{2 1}\in &
\left(
\frac{L_1}{L_1+L_2}\,,~~
\frac{L_1}{L_2}
\right)
\, , ~~~~~
\mathcal{T}_{1 2}\in 
\left(
\frac{L_2}{L_1+L_2}\,,~~
1
\right)
\, .
\end{align}

For $L_1=L_2$, the range of the tension is
\begin{align}
T\in
\left(
0\,,~~
\frac{2}{L_1}
\right)
\, ,
\end{align}
and the corresponding range of transmission coefficients are
\begin{align}
\mathcal{T}_{1 2}=\mathcal{T}_{2 1}\in 
\left(
\frac{1}{2}\,,~~
1
\right)
\, . 
\end{align}

\section{Entanglement entropy in holographic ICFT$_3$}
\label{app:b}

Consider the subsystem in the boundary, consisting of three regions  $-\sigma<x_1<0,-\sigma<x_2<0,-\sigma<x_2<0$. Suppose that the extremal surface $\Gamma$ intersects the junction brane at
\begin{align}
(t_\text{A},x_\text{A},u_\text{A})=
(0,z\sin\theta_\text{A},z\cos\theta_\text{A})
\, ,\ ~~~~ (\text{A}=1,2,3)\,,
\end{align}
and the endpoints of $\Gamma$ on the boundary are
\begin{align}
(t_\text{A},x_\text{A},u_\text{A})=
(0,-\sigma,\epsilon_{A}) \, ,\ ~~~~ (\text{A}=1,2,3)\,,
\end{align}
where the three UV cutoffs are defined by
\begin{align}
\epsilon_{A}\equiv\epsilon\cos\theta_\text{A} \, ,\ ~~~~ (\text{A}=1,2,3)
\,.\end{align}

The length functional is given by 
\begin{align}
\mathcal{A}(z)=L_a\sum_{\text{A}=1}^{3}
(\cos\theta_\text{A})\,\log
\frac{z^2+\sigma^2+2z\sigma\sin\theta_\text{A}}
{z\epsilon\cos^2\theta_\text{A}}  \, .
\end{align}
The extremal condition  $\frac{d\mathcal{A}}{dz}=0$ on the brane is
\begin{align}
(z^2-\sigma^2)\sum_{\text{A}=1}^{3}
\frac{\cos\theta_\text{A}}
{
z^2+\sigma^2+2z\sigma\sin\theta_\text{A}
}=0  \,  .
\end{align}
For $-\frac{\pi}{2} <\theta_\text{A}<\frac{\pi}{2}$, the solution is $z_{*}=\sigma$, giving the entanglement entropy
\begin{align}
\label{eq:entropy}
S_\text{E}=\frac{\mathcal{A}(\sigma)}{4G}=
\frac{L_a}{4G}
\sum_{\text{A}=1}^{3}
\cos\theta_\text{A}\log
\frac{2\sigma(1+\sin\theta_\text{A})}
{\epsilon\cos^2\theta_\text{A}}  \, .
\end{align}
The interface entropy is derived as 
\begin{align}
\begin{split}
S_\text{iE}=&S_\text{E}-
\frac{1}{4G}
\sum_{\text{A}=1}^{3}
L_\text{A}\log
\frac{2\sigma}{\epsilon\cos\theta_\text{A}}  \\
=&\frac{L_a}{4G}
\sum_{\text{A}=1}^{3}
\cos\theta_\text{A}\log
\frac{1+\sin\theta_\text{A}}{\cos\theta_\text{A}} \\
=&\frac{1}{4G}\sum_{\text{A}=1}^{3}
\rho_\text{A}^{*} \, ,
\end{split}
\end{align}
where $\rho_\text{A}^{*}$ is defined by
\begin{align}
\sin\theta_\text{A}=\tanh\frac{\rho_\text{A}^{*}}{L_\text{A}}  \, .
\end{align}

The $g$-function is then  
\be
\log g=4G\, S_\text{iE}=\sum_{\text{A}=1}^{3}
\rho_\text{A}^{*}\,.
\ee

This calculation naturally extends to  ICFT$_N$ for any $N\ge 2$,
\begin{align}
S_\text{iE}^{(N)}=
\frac{1}{4G}\sum_{\text{A}=1}^{N}
\rho_\text{A}^{*}  \, .
\end{align}
For $N=2$, $S_\text{iE}^{(2)}$ reduces to the known interface entropy in holographic ICFT in \cite{Anous:2022wqh}.

\end{document}